Comments arising from "Uncertainty in probabilistic genotyping of low template DNA: A case study comparing STRmix™ and TrueAllele®." Thompson WJ.

**Dear editor,**

Recently Thompson [1] reported his perception of a US federal Case in which he advised the prosecution. The case involved a Daubert hearing from which no ruling was given but did not proceed to trial. The evidence discussed was the interpretation of a low-level mixed DNA profile on a plastic bag containing drugs. A probabilistic genotyping (PG) software, STRmix™, had been used to interpret this mixture. This produced an *LR* that supported exclusion of the POI. A Daubert hearing was requested by the prosecution to suppress the DNA evidence. This is the first time, of which we are aware, that the prosecution has sought to suppress PG evidence.

There are sufficient doubts about the storage and handling of the bag that any source level conclusion is of limited value at the activity level. However, the source level is exactly what the discussion centered upon. Subjectively, using the subthreshold peaks, the data supports exclusion of the POI.

Thompson's paper covers a very broad range of topics. To retain some focus to our comments arising from this we only discuss here:

1. whether the analytical threshold (AT) should be varied in casework: "To learn more about the consequences of applying a threshold."
2. why TrueAllele® and STRmix™ give different answers, and
3. the number of contributors (NoC) should automatically be varied across the plausible range, and
4. the upper bound is the correct bound to report if the *LR* is below one, and
5. what do you do with multiple *LR*s.

**Varying the Analytical Threshold**

Thompson suggests in his paper that "*To learn more about the consequences of applying a threshold in this case, the prosecutor asked ESR to rerun STRmix™ in this case, using no threshold or a lower threshold.*" At the time Thompson suggested the rerun at lower AT because he could see peaks aligning with the accused.

The AT is usually set during validation of the capillary electrophoresis (CE) instrument and is not connected with the validation of any PG system. In this case it was set at 40 rfu on a ThermoFisher Applied Biosystems™ 3130/3130xl DNA Analyzer. ATs are usually set well above the level of noise. As such, subthreshold peaks are may often be true signal. The setting of an AT well above the noise is because the consequences of accepting an artifactual peak are very considerably negative and usually massively outweigh the slight improvement in true information gained by lowering the AT. This balancing requirement has been a consideration since the adoption of quantitive analytical techniques such as capillary electrophoresis. Once set the AT is usually not varied in casework and it should never be varied in an attempt to obtain more desirable *LR*s. The use of PG, which can consider the probability of drop-out for an allele, is preferred to lowering the AT to levels which increase detection of artifactual signals each of which has a chance of not being detected and removed at analysis. In fact the largest input file error we have noted for STRmix™ is the retention of an artifact in the input.

In the same section Thompson states: "*I think analysts who use STRmix™ to analyze DNA profiles of low-level DNA samples like the mixture in this case would be well advised to repeat the analysis with different analytic thresholds. The reason for repeated testing is not to find the result that is most incriminating or most exculpatory for the defendant; it is to test the robustness of the program by checking whether the findings hold up with different thresholds. If the LRs differ significantly for different thresholds, that finding should, of course, be reported.*" First we want to state that this behavior is absolutely not our recommendation. We can assure Thompson that the *LR* produced by STRmix™ will change if the AT is changed. As the AT is moved up the *LR* should, on average, tend to 1, that is those values above 1 should drop and those below should rise. As the AT is moved down the behavior is based entirely on whether correct or incorrect peak information is added to the inputs and whether the new information is concordant with the POI. Whether the change as the AT is lowered is beneficial or not cannot be learnt from the *LR*.

Thompson does appear to know this. In the next paragraph he states: *"The same is true for assessing the analytic threshold. Whether lowering the threshold increases the accuracy of the program by allowing it better to distinguish low-level contributors or reduces its accuracy by causing it to draw inferences from unreliable data can best be determined by testing accuracy with known-source samples of the type in question."* This quote from Thompson clearly described an experiment that could be done with mock samples but which is impossible in casework.

Hence, after the completion of Thompson's proposed experiment conducted in casework, the examiner is presented with a range of *LR* values but with no knowledge as to which of these should be recommended. Thompson suggests that these new values *"should, of course, be reported"* or perhaps he means that the fact that they vary should be reported. We are concerned that Thompson makes a recommendation to perform an experiment, presumably in every case, that cannot inform the decision making, that he himself appears to know is ineffective, and that he has not researched.

**Different answers from different software**

Thompson reports that *"For STRmix™, the reported likelihood ratio in favor of the non-contributor hypothesis was 24; for TrueAllele® it ranged from 1.2 million to 16.7 million, depending on the reference population."*

Later he opines that: *"By contrast, if the two programs produce widely different interpretations of the same evidence, then it suggests that PG findings are not always reproducible and raises concern that one or both programs may be untrustworthy."*

The term "*interpretation*" is used differently by different scientists. Based on context, it would seem that Thompson is using it to mean the *LR* magnitude and he notes that his view is that it is "*widely different.*"

There is an argument to be made that "interpretation" is best used to describe the expert's opinion and is separate from the PG outcome. In this process, which we apply, the scientists first examines the profile and compares the genotype of the POI to it. Potential outcomes of this process are that the data support inclusion of the POI, support exclusion of the POI, or that the data do not clearly support inclusion or exclusion. We are aware that other analysts run the PG analysis as a first, or an early, step.

In the case at hand, the data were edited at a (laboratory validated) AT of 40 rfu for STRmix™ and 10 rfu or 10 peaks for TrueAllele®. The two software both support exclusion in these analyses, so by this definition of "interpretation" STRmix™ and TrueAllele® are concordant. However, the inputs used by each software are different. It can hardly be a surprise that the *LR* magnitudes are different if the inputs differ. In fact it would be more surprising if the outputs were the same if the inputs differed.

Thompson does know this: *"Another issue worthy of consideration is the analyst's decision to apply a 40 rfu analytic threshold when running STRmix™. This decision meant that the data ultimately modeled by STRmix™ differed from the data modeled by TrueAllele® in subtle but potentially important ways that contributed to the discrepant results."* Given this statement by Thompson, with which we agree, we suggest that the central theme of his paper is not supported. For example the title: *"Uncertainty in probabilistic genotyping of low template DNA: A case study comparing STRmix™ and TrueAllele®"* appears to be based on the expectation that the results would be the same.

In subsequent correspondance Thompson has suggested that the results comparing TrueAllele® and STRmix™ reported below retrospectively vindicate exploratory AT changes. This has some credence if the goal was to test STRmix™ and TrueAllele®. But the goal is not that. It is to produce a statistic that the analysist feels is reasonable to present in court.

Thompson may not realise that in casework, practitioners operate almost exclusively following a standard operating procedure (SOP). Deviation from the laboratory SOP is permitted only for specific reasons, often requiring approval from the technical lead, and must be documented. Changes are not undertaken in casework for exploratory reasons. It is wrong to characterize the use of an AT of 40 rfu as a decision by the analyst. Any decision regarding AT was made much earlier at validation of the CE instrument and associated analysis software.

The observation of different outputs from different inputs would hardly need proof. Similarly different outputs from different PG models barely needs proof. The alignment of *LR*s obtained by lowering the AT for STRmix™ does not suggest that these values are correct nor that other values are incorrect. In fact we are not advanced much at all by this comparison.

TrueAllele® and STRmix™ V2.10 both have systems to deal with very low level peaks. STRmix™ V2.10 was only released in 2022 and we know of no lab routinely using a very low AT. Most of us are wary of very low peak heights. This feeling of discomfort is developed from a large body of experience noting the pernicious effects of artifacts that pass the analysis stage.

There are a number of studies now comparing different software (see for example [2-6]). The Riman et al [6] study is quoted by Thompson[1] but has accepted faults[2] noted by Riman et al

---

[1] Thompson also reprises the New York v Hillary case correctly quoting the PCAST addendum ([7]. Unfortunately PCAST misquote a newspaper. The reference to a newspaper is given clearly in PCAST and the addendum. The newspaper article is online and can easily be checked. It does not mention any TrueAllele or STRmix™ results. The STRmix™ result is known (LR ≈ 300,000). For TrueAllele we have no hard knowledge of the results but an email from Perlin ([8]) includes the statements:

"*The TrueAllele computer found no statistical support for a match between the fingernail scrapings from the left hand of the victim XXXXXXXXX and suspect XXXXXXXXX.*

"*There are suggestions of a possible match at some loci, i.e., positive log(LR) values. But there is measurable DNA degradation of the minor 5% component, with much apparent drop out and peak imbalance at a number of loci. The result is that the loci without match support outweigh the loci having such support.*

"*We did a number of computer runs (see below), including joint analysis of pairs of injections along with requests that assumed the known victim. At best, the loci cancelled each other out for a log(LR) sum near zero; most often the total was a solid negative match value. So while it is possible that XXXXXXXXX could have contributed his DNA to the sample, his contribution would be in the 1%-5% range, and is not statistically supported by the computer's analysis of this data.*"

If these findings do indeed differ the reason again is plausibly the much lower AT applied for TrueAllele. We do not extend this discussion further in order to avoid inflaming a highly charged topic.

[2] Riman et al note that they have misused EuroForMix and give a partial set of corrections. In Thompson's defense this revelation comes on page 19 of 30 and points to supplementary Tables S11 and S12, which are not part of the main body of the manuscript.

but more extensively in [4]. In general, the credible analyses show similar results with occasional notable discrepancies. We suggest that differences between software do exist and that this is proven. However, to be useful this knowledge needs be transformed into which, if any, of the software are credible and if so under what conditions. This is best undertaken by calibration [9] (see [10, 11] for some STRmix™ calibrations) but we do commend Thompson's locus by locus examination and the scoring against his subjective expectations. We consider this a vital part of the expert's interpretation process for all samples [12].

We record here our frustration with comparison studies. Charles Berger [13] gives a useful analogy. Imagine that two weather forecasters give two different forecasts for the weather. In such a situation we have obtained only the information that forecasts may differ. But in terms of predicting the weather we are no further advanced at all. Now let us further imagine that forecaster 1 has a 90% accuracy record of prediction for the last 1,000 days but that forecaster 2 has a 50% accuracy (nevermind that we have not defined accuracy). This now represents some useful information.

In closing this section we correct an error in Thompson's Table 1. He lists the weights given by STRmix™ and TrueAllele®. For STRmix™ these are $Pr(O|S_j)$ and, we believe, for TrueAllele® these are $Pr(S_j|O)$ where $O$ are the data and $S_j$ the genotype set proposal. These different probabilities are not directly comparable but can be interconverted.

**Varying the Number of Contributors (NoC) and Analytical Threshold (AT)**

The relevant profile (see Figure 1) shows one locus with three peaks labelled at AT = 40 rfu (note that the figure shows peaks down to 25rfu). All other loci show zero, one or two labelled peaks. It is important to note that when utilizing the apparent validated AT = 40 rfu, there is a single locus (D10S1248) with called peaks indicating more than a single contributor.

Our best interpretation of this mixture including the signals above 25 rfu is that this profile is more probable if it is a two person mixture than a single source profile. There is nothing exclusionary at the alleles that are above 40 rfu. However loci D3S1358, D8S1179, D2S441, D22S1045, and D10S1248 are missing apparent unshared alleles of the POI. These are the five lowest molecular weight loci in each dye channel.

There is a fair argument to be made that these missing alleles (at AT = 40) are sufficient for an interpretation of support for exclusion.

With AT = 25, there are four loci with obligate peaks from a second donor. As such this is "at least" a two person mixture but there is no evidence at all for more than two people even including the subthreshold peaks. This does not mean that it is impossible that it is a three person mixture, or indeed a four or five person mixture, just that there is no evidence for that.

There are 11 subthreshold alleles between 25 and 40 rfu. Of these four are condordant with the POI and 7 are not.

We re-ran STRmix™ V2.9 at an AT of 25 rfu as an academic experiment. This is about as low as we were prepared to go with V2.9 (V2.10 has the capability to go lower), and, it is also as low as the data have been reported in the public domain. This makes the inputs between TrueAllele® and STRmix™ more similar but still not the same. We used the NIST Caucasian database and generic parameters for STRmix™. Hence this is not casework quality analysis.

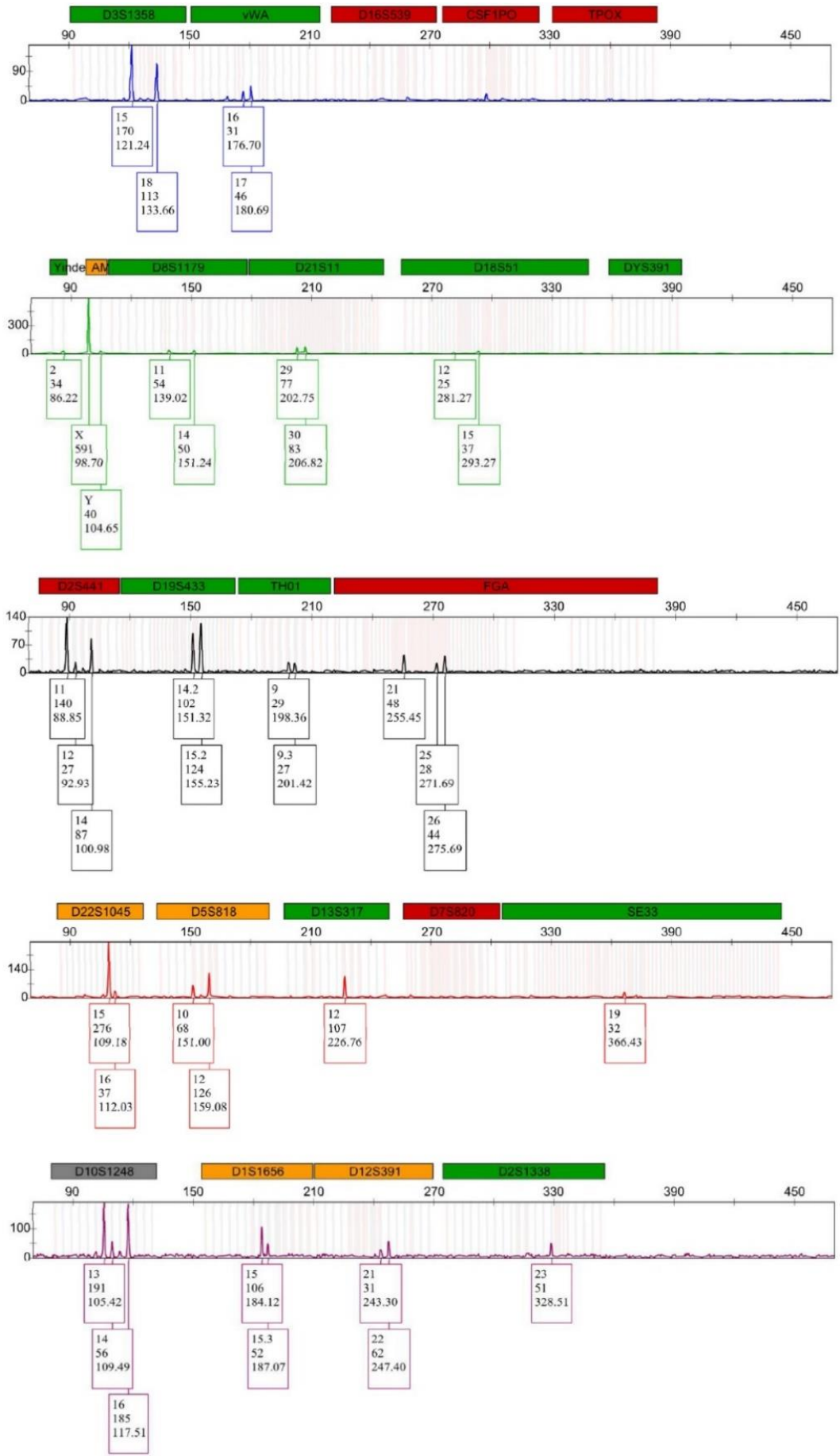

Figure 1. The electropherogram from the plastic bag (reproduced with permission from Pure Gold Forensics)

The overall *LR*s using the 25 rfu data (allowing for drop-in) for STRmix™ are $LR_{STRmix} = 8 \times 10^{-6}$ and $LR_{TrueAllele} = 6 \times 10^{-7}$. The *LR*s for D2S441 and D22S1045 provide the strongest exclusionary support at 2.2 x 10$^{-2}$ and 3.1 x 10$^{-2}$. Both of these have peaks attributable to a second donor which are non-concordant with the POI. No other locus *LR* was less than 0.31.

We know that STRmix™, at least, has low precision for *LR*s below one. The *LR*s are therefore not so different when the inputs are made more similar. We reiterate that we do not know if this new *LR* is more or less informative as that depends entirely on whether we have picked up additional real or artifactual signal.

Thompson opines: "*The DNA analyst, in this case, testified that he ran STRmix™ only under the assumption of two donors because he saw no evidence to support the presence of a third donor. ... Testing the robustness of the results by re-running the analysis under the assumption of three contributors would thus be a wise precaution in cases like this one.*"

We sought, and were granted, limited access to the raw data file for the sample in question to analyze the DNA trace at 10 rfu. We did this soley to see if there is anything at 10 rfu that is suggestive of three persons. This confirmed our impressions at 25 rfu that there is nothing that suggests the presence of a third contributor. As mentioned above, that does not mean that this cannot be three, four, or more contributors.

The general effect of adding a contributor is to increase low level false inclusionary support and reduce false exclusionary support [14]. The net effect of this is to move *LR*s closer to 1. To prove the point, we ran this sample in STRmix™ as three persons at 10 rfu. As expected, the *LR* returned was 0.25, which is a larger *LR* than the one reported in the case of 0.11. This was as expected. The addition of a phantom third contributor has reduced strongly exclusionary support toward an uninformative *LR*.

On balance, we view the outcome of adding an unneeded contributor as quite strongly negative since false inclusionary support more often has negative consequences. Certainly, the practice of doing it after finding exclusionary support gives a very strong appearance of attempting to shop for more palatable *LR*s. Adding this unnecessary contributor and moving the AT to 10 rfu would have almost "neutralized" the DNA evidence with respect to sub-source propositions.

The analyst is left, after this exercise, with a number of *LR*s and must find something to do with these. Biederman, Taroni, and Thompson gave, in 2011 [15], a valuable probabilistic approach that would weight the contributions to the *LR* seperately in the numerator and denominator based on prior probabilities for each number of contributors prior to observing the profile but assuming, respectively that the prosecution or the defense proposition is true.

We see almost no value in varying the inputs randomly across the plausible range and producing a range of outputs. Yet this is precisely what Thompson advocated.

There are at least two further approaches that exist that both assess the effect of uncertainty in the NoC. They both proceed by assuming that the NoC has a uniform prior distribution before examining the profile. These are:

1. The approach of Slooten and Caliebe [16] (hereafter S&C), and
2. The varNoC function in STRmix™ [17, 18].

To complete our analysis of NoC for this sample, we present the result of 3 persons in the numerator proposition and 2 persons in the denominator for both the S&C and varNoC methods. To begin, uncertainty in NoC exists in all profiles including single source profiles, not just the profile in this case.

Adapting to this case circumstance S&C give the *LR* considering uncertainty as

$$LR = LR_2 \Pr(NoC = 2 \mid E, H_a) + LR_3 \Pr(NoC = 3 \mid E, H_a)$$

Where $LR_2$ and $LR_3$ are the $LR$s assuming NoC = 2 or 3 respectively, and $E$ is the evidence profile, and $H_a$ is the alternative proposition that the POI is not a donor.

The $LR$s using STRmix™ V2.9, the generic parameters and the AT = 25 rfu input set are

$LR_2 = 8 \times 10^{-6}$ and $LR_3 = 5 \times 10^{-2}$

The $LR$ is therefore the weighted average of these two numbers. The reader may justifiably, at this point, note that both these results are support for an exclusion and that is likely to be all that they need. However we could also note that subjectively $\Pr(NoC = 2 \mid E, H_a)$ is much larger than $\Pr(NoC = 3 \mid E, H_a)$, that is $NoC = 2$ is much more probable given this profile than $NoC = 3$, and hence the overall $LR$ is nearer $LR_2 = 8 \times 10^{-6}$

The $LR$ obtained using varNoC is $LR_{2or3} = 6 \times 10^{-5}$ which is close to $LR_2$ as expected.

**Which bound if any to report if the *LR* is below 1?**

Thompson states: "… *when the results of PG analysis support the non-contributor hypothesis. In the range of possible LR values, the lower-bound estimate provides the strongest support for the noncontributor hypothesis. So reporting the lower-bound estimate, which is what the analyst did here, is the opposite of conservative. It provides the most extreme, rather than the most conservative, of the plausible LR values, which risks overstating the value of the DNA evidence for supporting the non-contributor hypothesis.*"

For the record, both bounds were available to the court and, perhaps more importantly, no trial occurred. Hence Thompson is referring to a report from the analyst. This includes only the lower bound. The software used was STRmix™ which does not make available the upper bound to users although we can obtain that at the STRmix™ group. We are considering whether we should add the functionality to give both bounds.

We are not aware of any discussion on which bound to report from SWGDAM, OSAC, or elsewhere, and we certainly do not think that any aspersion should be cast upon the analyst. SWGDAM [19] appear to recommend an upper bound for an *LR* in support of exclusion, such as 0.01, but there is no discussion of reporting an *LR* that is more conservative toward the POI (lower bound) or toward the evidence (upper bound) for a distribution of *LR*s that are all below 1. However, it may be suitable for this matter to be considered by, for example, SWGDAM.

Again, simply for the record, we note a very valid view, with which we have much sympathy, that quoting a bound on the *LR* may be wrong in the first place and that the best assignment of the *LR* is the one that should be reported [20-22]. An additional benefit of reporting, for example, the *average LR* is that such a term may remind stakeholders that there are a range of *LR*s rather than "the" *LR* that is reported. We are adherents to this view despite providing tools to the community to obtain bounds.

**What should be done with multiple *LR*s?**

Let us imagine that we learnt *"more about the consequences of applying a threshold in this case… [by rerunning] STRmix™ in this case, using no threshold or a lower threshold"* and tested *"the robustness of the results by re-running the analysis under the assumption of three contributors"* we would now have at least four *LR*s. If we then add supplying the court with the upper and lower bound and the best assignment, then there will be 12 numbers which could be presented to the court. We would suggest that the analyst now must do something. This could be to deliver all *LR*s to the court but recommend one. But if you are going to recommend one why not deliver that and somehow disclose that there are others that you have done for some reason but that you do not recommend? Whatever the analyst now does, we would suggest, will be used by the adversarial system. In any case we strongly recommend against *LR* shopping, hence the decision of which one to recommend must not be made based on the result.

Taroni [23] gives us the insightful comment: "*You write in a report what you are able to justify respecting your standards, tests, scientific knowledge.*"

Recall that this comparison subjectively supports an exclusion of the POI. No *LR* at all was really required. This profile is from the outside of a plastic bag. There are doubts about the storage and handling of this item. We observe that the discussions for the Daubert (that did happen) and trial (that never happened), Thompson's paper and our reply represent a lot of analysis of a very low information content profile. We would feel that this was not warranted.

We close with a reminder from a prior publication [24] where Thompson criticizes forensic DNA analysts for "*overestimat[ing] likelihood ratios*" by "*shifting the purported criteria for a 'match' or 'inclusion' after the profile of a suspect becomes known—a process analogous to the well-known Texas sharpshooter fallacy.*" Some readers might find it hard to accept Thompson's statement [1] of "*While I certainly agree that it is inappropriate, and even unethical, to raise and lower analytic thresholds in a goal-directed effort to achieve a certain result in litigation, I believe the analysis requested by the prosecutor would have yielded useful information.*" Such readers might think this is nothing more than an endorsement of the Texas sharpshooter fallacy updated for modern probabilistic genotyping.

Tim Kalafut
Department of Forensic Science, College of Criminal Justice, Sam Houston State University, Huntsville, TX, USA

James Curran
Department of Statistics, University of Auckland, Private Bag 92019, Auckland 1142, New Zealand

Michael D. Coble
Center for Human Identification, Department of Microbiology, Immunology, and Genetics, University of North Texas Health Science Center, 3500 Camp Bowie Blvd., Fort Worth, TX 76107, USA

John Buckleton
Department of Statistics, University of Auckland, Private Bag 92019, Auckland 1142, New Zealand
Institute of Environmental Science and Research Limited, Private Bag 92021, Auckland 1142, New Zealand; john.buckleton@esr.cri.nz